\documentclass[12pt]{article}
\begin{document}

\centerline{\Large Krichever I.M.\footnote{Krichever I.M.,
Colunbia University, New York NY 10027, USA and Landau Institute for
Theoretical Physics, Kosygina str 2, 117940 Moscow Russia, e-mail
krichev@math.columbia.edu.
Research supported in part by the NSF Grant DMS-9800257.}, Novikov S.P.
\footnote{
Novikov S.P., University of Maryland, College Park MD 20742-2431 and Landau Institute
for Theoretical Physics, Kosygina str 2, 117940, Moscow Russia,
e-mail novikov@ipst.umd.edu. Research supported in part by the NSF Grant DMS-9704613}}
\vspace{0.1cm}
\centerline{\bf Holomorphic Bundles and Difference Scalar Operators:}
\centerline{\bf One-point constructions}

\vspace{0.2cm}
Let us consider a nonsingular algebraic curve $\Gamma$ with marked point
$P_0=\infty$ and local coordinate $z=k^{-1}, z(P_0)=0$ like in \cite{kn}.
Let $A=A(\Gamma,P_0)$ be a ring of algebraic functions with  only pole
at the point $P_0$. Our inverse spectral data consist of the set of points
$\gamma_s,s=1,\ldots,lg$, where $l$ is ''rank'' of our construction (see below)
and $g$--genus of the curve $\Gamma$, parameters $\alpha_{sj},s=1,\ldots,lg,
j=1,\ldots,l-1$, and matrix-valued $l\times l$--functions
$\chi^{(0)}_n(k)$ with only nonzero elements $\chi_n^{(0)p,p+1}=1,p\leq l-1,
\chi_n^{(0)lq}=a_{qn},q=1,\ldots,l$. Here $a_{qn}$ are polinomials in
the variable $k$ depending on $n$.

\newtheorem{theo}{Theorem}

\begin{theo} For any fixed vector $\eta_0$, and a generic data
there exists and unique the ''Baker-Akhiezer'' $l$-vector-function
$\psi_n(P),P\in\Gamma$ that is meromorphic outside  the point $P_0$
with first order poles at the points $\gamma_s$  such that
$$res_{\gamma_s}\psi^{q+1}=\alpha_{sq}res_{\gamma_s}\psi^1,s=1.\ldots,lg$$
In the neighborhood of the point $P_0$ this function $\psi$
has an asymptotic behavior
\begin{equation}\label{1}
\psi=[\eta_0+\sum_{p\geq 1}\eta_{pn}k^{-p}]\Psi_n^{(0)},
\end{equation}
where $\Psi_n^{(0)}$ is defined by the equation
$\Psi^{(0)}_{n+1}=\chi^{(0)}_n\Psi^{(0)}_n$.
\end{theo}

\begin{theo} Let  matrix $\chi_n^{(0)}$ be such that
$a_{jn}=k+v_{jn}^{(0)}$, all other elements $a_q=v_{qn}^{(0)},q\neq j$
do not depend on $k$, and let $\eta_0$ be a vector with the cooordinates
$\eta_0^i=\delta^{ij}$. Then for every function $f\in A(\Gamma,P_0)$
with pole of order $\tau$ there exists a unique difference
operator $L_f$
$$L_f=\sum_{-M}^{+N}u_{pn}T^p$$
where $T\psi_n=\psi_{n+1}, M=\tau(j-1),N=\tau(l-j+1)$ such that
the Baker-Akhiezer vector function defined by our data satisfies
the equation
$$L_f\psi=f\psi.$$
\end{theo}

{\bf Remark}. These statements are descrete analogs of
the results obtained in \cite{kn,kr1} which give a construction of
higher rank commuting differential operators and correspodning solutions
of the KP equation. Let us remind that $(\alpha,\gamma)$ is exactly the set of
 ''Tyurin Parameters''
characterizing the stable framed holomorphic $l$--bundles ${\cal E}$
over $\Gamma$ where $c_1(\det{\cal E})=lg$. All previously
known constructions of the commuting difference
operators were of the rank 1. They always required exactly two infinite
marked points $\infty_{\pm}\in\Gamma$. Let us point out also
that in our one-point rank $l>1$ construction presented here,
the symmetric operators $M=N$ can be obtained
for the even rank $l=2(j-1)$ only.

Following the idea of \cite{kn}  multiparametric
Baker-Akhiezer vector-function can be defined through the same inverse data
$\Gamma,P_0,z=k^{-1},\gamma,\alpha,\chi^{(0)}$, but for  every new time
variable $t_p$ we have to use corresponding matrix $M^{(0p)}_n(t)$,
where $t=(t_1,t_2,...),p=1,2,...$. The initial matrix $\Psi^{(0)}$ can be
constructed using the integrable system (hierarhy):
$$\Psi^{(0)}_{n+1}=\chi^{(0)}_n\Psi^{(0)}_n, \ \
\Psi^{(0)}_{t_p}=M_n^{(0p)}\Psi^{(0)}$$
For example let us take matrix functions $M_n^{0+}$ and $M_n^{0-}$
of the form
$$
M_n^{0,1+}=\left(\begin{array}{ccccc} w_{1n}& 1&0&..&..\\
                                    0& w_{2n}&1&..&..\\
                                    ..&..&..&..&..\\
                                    0& 0&...&w_{l-1,n}&1\\
                                  a_{1n}&a_{2n}&...&a_{l-1,n}&a_{ln}+w_{ln}
\end{array}\right)
$$

$$
M_n^{0,1-}=\left(\begin{array}{ccccc} c_{1n}b_{2n}&c_{1n}b_{3n}&..&c_{1n}b_{ln}&c_{1n}\\
                                    c_{2n}&0&0&..&..\\
                                    ..&..&..&..&..\\
                                    0& 0&c_{l-1,n}&0&0\\
                                     0&0&...&c_{ln}&0
\end{array}\right)
$$
where $c_{qn},w_{qn}$ do not depend on $k$, the functions $a_{qn}$ are the same
as in theorem 2, and $b_{qn}=a_{1,n-1}^{-1}a_{qn}$.

\begin{theo}
For any $l\geq 2$ we can choose  matrices $M^{(0p)},p\geq 1$, such that
the Baker-Akhiezer vector-function $\psi_n(t),t=(t_1^+,t_1^-,t_2,...)$
corresponding to any generic data
$\Gamma,P_0,z=k^{-1},\gamma,\alpha, \chi^{(0)}$
defines with the help of the formula
$$\phi_n(t)=y_n+\ln(1+\eta_{1n}^{j-1})$$
a solution of 2D Toda lattice hierarchy.
Here $y_n$ are such that $c_n=\exp(y_n-y_{n-1})$, and $\eta_{1n}^{j-1}$
is the $(j-1)$-th component of the vector $\eta_{1n}$ in the expansion
(\ref{1}).
\end{theo}
{\bf Example}. Let $g=1,l=2, a_1=-c^{(0)}_{n+1},a_2=k-v^{(0)}_{n+1}$,
the data $(\Gamma,k^{-1}=z,P_0=0,\gamma_1,\gamma_2,\alpha_1,\alpha_2)$
 and the Weierstrass function $\wp(z)\sim k^2$ are given.
For  the Baker-Akhiezer
matrix $\hat{\Psi}_n$ whose rows are $\psi_n,\psi_{n+1}$ we have
$\hat{\Psi}_{n+1}=\chi_n\hat{\Psi}_n$ where the rows of $\chi$ are
  following:
$$\chi_n^1=(0,1),\ \ \chi^2_n=(-c_{n+1},k-v_{n+1})+O(k^{-1})$$
Poles of the matrix-function $\chi_n$ are located in the points
 $\gamma_{sn}$ where $\gamma_{s0}=\gamma_s,s=1,2$.
Zeroes of the function $\det\chi_n$ are located at the points
 $\gamma_{s,n+1}$. We have following relations
$$\alpha_{sn}res_{\gamma_{sn}}\chi^{i1}=res_{\gamma_{sn}}\chi^{i2},i=1,2$$
and $\alpha_{s,n+1}=-\chi^{22}(\gamma_{s,n+1})$.
 Our curve $\Gamma$ is realized as a complex plane with the coordinate $z$
factorized by the lattice of periods. The sum $\gamma_{2n}+\gamma_{1n}=c$
does not depend on $n$. For every function $f\in A(\Gamma,0)$ we can compute the commuting
operators $L_f$ given by the theorem 2. For the Weierstrass function
 $f=\lambda=\wp(z)$ and $c=0$ we have  symmetrizable fourth order operator
$L_{\lambda}$ given by the formulas:
$$\lambda\psi_n=L_{\lambda}\psi_n=[(L_2)^2+u_n]\psi_n,L_2\psi_n=
\psi_{n+1}+v_n\psi_n+c_n\psi_{n-1}$$
$$u_n=-[\wp(\gamma_{n-1})+\wp(\gamma_{n-2})]+b_{n-1}+b_{n-2},\wp(z)=-\zeta'(z)$$
$$b_n=2\wp'(\gamma_n)[\wp((\gamma_{n+1}+\gamma_n)-\wp(\gamma_{n+1}-\gamma_n)]
[\wp'(\gamma_{n+1}+\gamma_n)-\wp'(\gamma_{n+1}-\gamma_n)]^{-1}$$
$$c_n=(\alpha_{1n}-\alpha_{2n})^{-1}[\zeta(\gamma_{n+1}-\gamma_n)
-\zeta(\gamma_{n+1}+\gamma_n)+2\zeta(\gamma_n)]$$
Here we have $\gamma_n=\gamma_{1n}$ by definition, $\gamma_n$ and $v_n$ are
the arbitrary functions. The functions $\alpha_{in},i=1,2$ can be found from
the formulas:
 $$\alpha_{i,n+1}=-v_{n+1}+(-1)^{i-1}\zeta(\gamma_{n+1})+(-1)^{i-1}
\alpha_{1n}(\alpha_{1n}-\alpha_{2n })^{-1}
\zeta(\gamma_{n+1}-\gamma_n)+$$
$$+(-1)^{i-1}\alpha_{2n}
(\alpha_{1n}-\alpha_{2n})^{-1}
\zeta(\gamma_{n+1}+\gamma_n)$$
Consider now  time dynamics in the variable $t=t_1^+$
such that $M^{(0,1+)}_n(t)=\chi^{(0)}_n(t)+diag(v^{(0)}_n,v^{(0)}_{n+1})$.
For the Baker-Akhiezer matrix $\hat{\Psi}_n$ we have
$$\partial_t\hat{\Psi}_n=M_n\hat{\Psi}_n,\hat{\Psi}_{n+1}=\chi_n\hat{\Psi}_n$$
$$M_n=\chi_n+diag(v_n,v_{n+1})+O(k^{-1})$$
>From the compatibility conditions in the variables $n,t$ we are coming to
the nonlinear system for $c_n(t),v_n(t))$:
$$\partial_tc_{n+1}=c_{n+1}(v_{n+1}-v_m),\partial_tv_{n+1}=c_{n+2}-c_{n+1}+
\kappa_{n+1}-\kappa_n$$
Here we have $\chi^{22}_n=k-v_n+\kappa_nk^{-1}+O(k^{-2})$.
This system is a difference analog of the so-called ''Krichever-Novikov''
system \cite{kn}. It parametrizes some family of the ''rank 2''
solutions to the
2D Toda Lattice instead of KP. No problem to compute
effectively the coefficient
$\kappa$ using dynamics of the Tyurin parameters.

{\bf Remark}. In our next note we present two-(and more)--point constructions
of the rank $l\geq 2$ containing some very interesting new phenomena.

\end{document}